\documentclass[conference]{IEEEtran}
\usepackage{graphicx}
\usepackage{amsmath,amssymb,color,cite}
\usepackage{algorithmic,algorithm}

\usepackage{pgf,tikz}
\usetikzlibrary{arrows}
\definecolor{zzttqq}{rgb}{1,0,0}
\definecolor{qqqqff}{rgb}{0,0,1}
\definecolor{cqcqcq}{rgb}{1,1,1}

 \newtheorem{theorem}{Theorem}[section]

 \newtheorem{proposition}[theorem]{Proposition}

 \newtheorem{remark}[theorem]{Remark}

\begin{document}

\title{Incentive Design for Direct Load Control Programs}

\author{Mahnoosh Alizadeh$^\dagger$, Yuanzhang Xiao$^{\star}$, Anna Scaglione$^\dagger$, and Mihaela van der Schaar$^{\star}$\\
$^\dagger$University of California Davis~~~ $^\star$University of California Los Angeles
}


\maketitle

\begin{abstract}We study the problem of optimal  incentive design for voluntary participation of electricity customers in a Direct Load Scheduling (DLS) program, a new form of Direct Load Control (DLC) based on a three way communication protocol between customers, embedded controls in flexible appliances, and the central entity in charge of the program. Participation decisions are made in real-time on an event-based basis, with every customer that needs to use a flexible  appliance considering whether to join the program given current  incentives.
Customers have different interpretations of the level of risk associated with committing to pass over the control over the  consumption schedule of their devices to an operator, and these risk levels are only privately known. The operator maximizes his expected profit of operating the DLS program by posting the right participation incentives for different appliance types, in a publicly available and dynamically updated table. Customers are then faced with the dynamic decision making problem of whether to take the incentives and participate or not. We  define  an optimization framework to determine the profit-maximizing incentives for the operator. In doing so, we also investigate the  utility that the  operator expects to gain from recruiting different types of devices. These utilities  also provide an upper-bound on the benefits that can be attained from any type of demand response program.
\end{abstract}

\section{Introduction}\label{intro}
With the  lack of utility-scale storage options in the power grid, the need to make  electricity  demand active is becoming more pressing each day. In the research community, the most favored option to make this vision happen is real-time pricing (RTP).  Optimal real-time prices, if calculated correctly, would maximize the social surplus. However, there are several barriers that currently hinder the realization of this vision: 1) End-use customers need to have certainty in prices for a certain look-ahead horizon to plan consumption. With a wide-spread integration of renewables, calculating reliable  clearing prices hours ahead of operation is challenging; 2) the lack of concrete models for the consumption flexibility of electricity consumers in today's market, specifically with the granularity needed to allow high penetration of renewables, and the time-inhomogeneity of these models due to variable  appliance arrival and flexibility patterns; 3) the strict reliability requirements of  power grid operations,  allowing small error margins in price design. A line of iterative methods that actively ask for the customers' collective response to price signals are being proposed to address these problems \cite{rtp,rtp2}.

One the opposite side of the spectrum, Direct Load Control (DLC) has proven to be a popular type of demand control for grid operators, mostly due to the reliable and predictable nature of the demand's response to control signals.
  Currently, customers providing DLC services to the power grid sign long-term contracts that allow the  provider to cut off the supply of electricity to some of their appliances (e.g., air conditioners and pumps) during occasional contingencies. 
Commonly, DLC strategies take customer participation as a given, assuming small fixed ex-ante monetary payments for all participants, regardless of the level of service they provide or the discomfort they experience. However, participation in a DLC program presents risks for the customers and a rational intelligent customer 
would not provide this service extensively without appropriate financial incentives.  In this work, we take a first step in addressing these inherent economic problems for designing day-to-day DLC incentives.

To avoid confusion with currently employed DLC programs, in which the individual consumption of appliances are not observable, and feedback control strategies are used \cite{feed1}, we will refer to our proposed program as Direct Load Scheduling (DLS). DLS does not merely cut off the electricity supply of appliances. Rather, a control center can optimally {\it plan} the consumption within consumer-specified laxity limits, e.g., scheduling the charge of an Electric Vehicle (EV) by a deadline. Contrary to common DLC practice, customers provide the DLS authority with an explicit expression of the service they need in an online fashion, leaving no uncertainty in how the demand responds to control signals. Appliances are recruited only on an event-based basis, i.e., every time they have to perform a task. Previous works have proposed various cost-minimizing scheduling algorithms for DLS, specifically for EVs, e.g. \cite{evdlc,evdlc2,evdlc3,evdlc4,evdlc5,evdlc6}. The possibility of using Vehicle-to-Grid (V2G) services for regulation and grid support has also been studied, e.g. \cite{v2gev,v2gev2}. However, charge interruption and V2G could decrease EV battery life, delay full charge and present risks to the customer, and violate customer privacy. Thus, proper economic incentives for participation should be studied.

We assume that the DLS program is run by the same entity that provides electricity to customers, hereafter referred to as the {\it aggregator}.  The aggregator is essentially an energy {\it trader}. Following current practice, we assume that the aggregator buys electricity at 
time-varying wholesale prices from the energy market and on sells this energy to end-use customers at flat rates, without being able to deny any electricity service requests. Thus, it essentially acts as intermediary node that   shields the end-use 
customers from wholesale price fluctuations. With no demand flexibility, the
aggregator has no control over the profitability of his venture in the short run. Rather, his profit is
determined by 1) wholesale prices; 2) the predetermined billing tariffs, which are flat, regulated, and 
change very slowly; 3) the consumption behavior of customers, which is out 
of control of the aggregator.
To overcome this issue, the aggregator runs a DLS program and pays customers in return for directly scheduling their appliances, i.e., it effectively buys flexibility from customers  while keeping them on flat rates.  This paper focuses on the  design of a market for trading flexibility between a single aggregator and a population of randomly arriving appliances. We model this market as a monopoly, in the sense that each customer has access to only one aggregator and is a price-taker. Competition between aggregators is left to future work.

%
%

%

\section{Model}

We consider a community with a large population of customers owning controllable appliances, and one   aggregator. The basic observation of this work is that directly controlling various types of appliances has different utilities for the aggregator. This utility is a function of how flexible and how high the electricity consumption of the device is, and of the dynamic state of the grid at the time the appliance is used. Thus, it is best if the incentive that the aggregator pays to customers to recruit their appliances can vary dynamically with {\it time and appliance type}\footnote{ 
 While, in theory, the incentive could vary across different customers offering the same service, here we assume that we want the  incentives to be perceived as fair and not violate consumer privacy limits.}. To allow for this ,we assume that these  incentive are posted by the aggregator in dynamically updated and publicly available tables for all customers, which can be thought of as {\it menus} listing the different incentives customers can receive. Customers planning to use controllable appliances would then respond to these posted incentives by deciding whether they want to participate or not, and how much laxity they wish to offer the aggregator. 

We would like to point out that having dynamically changing menus does not necessarily mean that the incentives have to be re-designed every hour of every day. In fact, incentives will exhibit similar daily or weekly cycles as market prices.

We define this market under the following assumptions: \\
\textbf{Assumption 1}: The aggregator's revenue from recruiting  each appliance is additive and independent of other appliances' participation;\\
\textbf{Assumption 2}: A customer's initial choice to use an appliance  is perfectly inelastic and not affected by   incentives; \\
\textbf{Assumption 3}:  The aggregator's load does not affect the wholesale market clearing prices;\\
\textbf{Assumption 4}: The aggregator has access to ex-ante forecasts of expected wholesale prices. 

Assumptions 3 and 4 simplify the expression of the aggregator's utility when recruiting appliances in the DLS program.
Next, we look at how customers respond to incentives.

\subsection{Individual Appliance DLS Commitment Problem}\label{custmodel}

We model the electricity consumption of  customers using tasks that dynamically arrive, receive service, and depart at discrete time epochs indexed by $t=\{1,2,3,\ldots\}$.  An arrival event corresponds to the earliest time at which it is possible for an appliance to start its job.  Each task, indexed by $i$, has a so-called characteristic vector $\mathbf{v}_{i} \in  \mathcal{C}$. The elements of $\mathbf{v}_{i}$ fully describe the nature of the task, which could simply be the charge duration and rate for an EV,  the desired temperature for a thermostatically controlled load (TCL). We further assume that $\mathbf{v}_{i}$ can only be chosen from a finite codebook $\mathcal{C}  = \{\mathbf{c}_1,\mathbf{c}_2,\ldots,\mathbf{c}_Q\}$, designed to achieve a bounded and controllable load modeling error. This allows us to cluster similar energy requests in a finite number of classes, indexed by 
$$q_i  \in \mathcal{Q} = \{1,2,3,\ldots, Q\},$$
 which will help to highly reduce the computational effort of our algorithm. Thus, the consumption characteristics of task $i$ are uniquely defined by its cluster index $q_i$.

The strategy set is defined by the laxity limits the customer commits to provide to the DLS program  when recruited, described by the index 
$$m_i \in  \mathcal{M}_{q_i} = \{0,1,2,\ldots,M_{q_i}\},$$
  and referred to as the appliance's {\it mode}. Mode $m=0$ corresponds to {\it no laxity}, i.e., the customer will not participate in the program. These laxity limits could, for example, correspond to the slack for an EV charge, or the  width of the comfort band for a TCL.  The appliance embedded controller is queried by the customer application, which maps the physical state of the appliance onto the set of modes that are available for the customer to choose from and, possibly, an indication of what is the flat rate cost with no laxity.

The customer then chooses the mode after observing the incentive menu. 
We denote by  $I^{t}_q(m)$  the incentive the customer could receive for releasing the control of an appliance in cluster $q$ in mode $m$ at time $t$, with $I^{t}_q(0) = 0$. We  denote the vector containing all (non-trivial) incentives available to appliances in cluster $q$ under modes $m\geq 1$ as
$$\mathbf{I}^{t}_q = [I^{t}_q(1),I^{t}_q(2),\ldots,I^{t}_q(M_q)]^T.$$

 We expect the customers in charge of making commitment decisions to be only {\it boundedly rational}, i.e.,  they are likely to only spend a limited amount of effort in considering the economic utility of participation in the program or updating their home energy management system's parameters. However, in the rest of this section, we introduce an analytical model which is valid for rational customers. We do so with the disclaimer that this model is not a necessary element of our design and we present  it solely to provide some intuition into the nature of this decision making problem.

\textbf{Design constraint 1} (Diminishing Payoffs): Note that if an appliance does not join the DLS program right after its arrival time, it will {\it lose} some laxity. However, if consumers have reasons to believe that taking some risk and waiting to participate in the DLS program later could increase their expected payoff, even though the amount of laxity they can offer decreases, they will do so. To avoid this situation,  we assume that the incentives will be designed such that the customer's payoff for the same level of risk (e.g., same deadline) will be monotonically non-increasing in time. For example, if  we assign a separate mode $m$ every time the laxity offered by a deferrable load is increased by one unit, we require that $$I^{t}_q(m) \leq I^{{t}-1}_q(m+1).$$

\textit{A Rational Customer Model}: The utility gained by the customer from operating task $i$ at time $t$ in each mode is defined as a function 
$$V^t_i(.): \mathcal{M}_{q_i} \rightarrow \mathbb{R},$$ which includes three terms:
1) The incentive  $I^t_{q_i}(m_i)$ available for mode $m_i$; 2) The commitment risk (disutility) $R^t_{i}(m_i)$ associated with agreeing to receive service under mode $m_{i}$, modeled through a privately known function
$$R^t_{i}(.): \mathcal{M}_{q_i} \rightarrow \mathbb{R}^+,$$
with $R^t_i(0) = 0$, i.e., there is no risk when the customer decides not to participate in the DLS program. Without loss of generality, we order the modes from low to high risk. This would result in monotonically non-decreasing individual risk functions $R^t_i(m_i)$; 3) the utility of receiving the standard service of using electricity and finishing a job. This term is a constant,  since we assume it will eventually  happen for every request, either through the DLS program under some mode $m_i \geq 1$, or through the standard service model of the power grid ($m_i = 0$). The disutility of not receiving this standard service in case of an emergency departure is captured through the risk term. Thus, we eliminate this term from the customer's decision making model.

Consequently, assuming that the risk function are chosen such that they have the same unit as the monetary incentive,  the customer's utility is quasi-linear and is given by
$$V^t_i(m_{i})= I^t_{q_i}(m_i)- R^t_{i}(m_{i}).$$

Upon receiving the incentive information $I^t_{q_i}(m_i)$ from the aggregator, the customer would solve the following optimization to determine the best operating mode of appliance $i$:
\begin{eqnarray} \label{argmaxstep} \max_{m_{i} \in \mathcal{M}_{q_i}}&~&V^t_i(m_{i}). \label{custopt} \end{eqnarray}
The customer will not participate in the DLS program ($m_i=0$) if none of the $I^t_{q_i}(m_i)$'s for $m_i\geq 1$ are at least marginally higher than the risk $ R^t_{i}(m_{i})$. Otherwise, the customer would pick the mode such that the margin between $I^t_{q_i}(m_i)$ and $ R^t_{i}(m_{i})$ is highest (highest residual worth). Ties are broken uniformly at random.


Even if we assume that this model perfectly describes the decision making procedure of all customers, since $R^t_{i}(.)$ is only privately known, the aggregator cannot predict the outcome of  \eqref{custopt} deterministically. Thus, to determine the
incentives $I^t_{q}(m_{i})$, the aggregator is faced with 
an optimization problem with incomplete information.  The goal of the aggregator would be to maximize its expected profit, given aggregate statistics about the population's response to incentives.

\subsection{The Aggregator Problem}

In order to recruit  directly controllable appliances, the aggregator needs to design appropriate incentives for every possible mode in all possible clusters, i.e., $I^t_q(m), \forall q \in \mathcal{Q}, \forall m \in \mathcal{M}_q$.
The LSE participates in the electricity wholesale market   run on an hourly basis, where it buys energy to serve its load, directly controllable or not. We will elaborate more on the nature of this market interaction later. For now, we take the the expected profit that the aggregator can make, through wholesale market transactions, by directly controlling an individual appliance from cluster $q$ in mode $m$ as known. For a recruitment of cluster $q$ in mode $m$ at time $t$, we denote this utility by   $U^t_q(m)$.  The utility of not recruiting an appliance is zero, i.e., $U^t_q(0) = 0$. As per Assumption 1, we define this recruitment utility as additive and independent, rendering the incentive design problem separable for individual appliances.

 Since recruiting an appliance from cluster $q$ in mode $m$ at time $t$ comes at a cost equal to $I^t_{q}(m)$, the net revenue of the aggregator from this recruitment, denoted by $N^t_q(m)$, is
\begin{equation}N^t_q(m) = U^t_q(m) -  I^t_{q}(m).\end{equation}
 
However, note that the mode $m$ is chosen by the customer after seeing the incentives $I^t_{q}(m)$, through \eqref{argmaxstep}. Since only statistical information on the customers response strategy is available to the aggregator, only the {\it expected net revenue} can be maximized. Denote the event that any customer in cluster $q$ picks mode $m$ as  $E_{q,m}(\mathbf{I}^t_q)$. This event happens  if:
\begin{itemize}
\item Individual rationality constraint (\textbf{IR}):
$$ I^t_{q}(m)- R^t_{i}(m) \geq  0,$$ 
\item Incentive compatibility constraints (\textbf{IC}):
$$ I^t_{q}(m)- R^t_{i}(m) \geq   I^t_{q}(m')- R^t_{i}(m') ,~~\forall m' \in \mathcal{M}_q.$$
\end{itemize}
Consequently, the expected net revenue of recruiting an appliance of cluster $q$, simply denoted by $N^t_q$, is given by
\begin{equation} \label{profitexp} N^t_q =  \sum_{m \in \mathcal{M}_q} P(E_{q,m}(\mathbf{I}^{t}_q ))( U^t_q(m) -  I^t_{q}(m)),\end{equation}
which the aggregator would like to maximize; i.e. the aggregator would want to design the incentives  as follows:
\begin{equation} \label{profitmax} \max\limits_{ \mathbf{I}^{t}_q}~\sum_t N^t_q.\end{equation}
The summation over time is required to find the optimal incentives in the presence of the diminishing payoff design constraint. The reader could envision that for a real-time implementation, the optimization \eqref{profitmax}  could be solved over a receding horizon.

In order to solve \eqref{profitmax}, it is essential to have an understanding of how $P(E_{q,m}(\mathbf{I}^{t}_q ))$ changes with the incentives $\mathbf{I}^{t}_q $. Next, we propose two different views for approaching this problem.

\subsubsection{Bayesian Approach}
Here we assume that the aggregator has  access to statistically learned prior knowledge on the risk levels of customers in different clusters. With this view, the incentive design problem would be similar in nature to optimal Bayesian unit-demand pricing, given that the customers' risk levels (valuations) for different modes (items) cannot be assumed to be drawn from independent distributions.
The risk levels that a customer perceives for committing to the program with a single appliance under different modes are correlated. For example, in the case of EVs or any other deferrable loads, the mode corresponds to the amount of laxity that accompanies the request. Clearly, offering a higher consumption laxity entails an additional risk over that of offering a lower laxity, and these variables cannot be considered independent.

We assume that the aggregator parameterizes the risk function of each appliance $i$ in cluster $q$ according to:
\begin{equation} \label{typeeq}R^t_i(m) = \gamma_i r_q^t(m),\end{equation}
where $\gamma_i$ is a task-specific non-negative continuous random variable and represents the {\it type} of an individual task $i$, and the variables $r_q^t(m)$ are deterministic and shared by all appliances in the same cluster $q$, with $r_q^t(0) = 0$, and can model the average attitude of  population towards risk. To ensure that this is a realistic parameterization of the risk function, the aggregator could suggest this specific structure as default to consumers when they pick their risk functions.

 We assume that the aggregator has access to  statistical priors for the type $\gamma_i$, and we denote by $F_{\gamma}^{q}(g)$ the cumulative distribution of $\gamma_i$ for appliances in cluster $q$. With this new notation, $P(E_{q,m}(\mathbf{I}^{t}_q ))$ for all $m\geq 1$ is the probability of the following event:
\begin{align} &\mbox{\textbf{IR}} - \gamma_i  \leq \frac{I^t_{q}(m)}{r_q^t(m)} = \frac{I^t_{q}(m)- I_{q}(0)}{r_q^t(m) - r_q^t(0)},\label{ir}\\
 &\mbox{\textbf{IC}1} - \gamma_i  \leq \frac{I^t_{q}(m)- I^t_{q}(m')}{r_q^t(m) - r_q^t(m')} ,~~\forall 1\leq m'<m \label{ic1}\\
 &\mbox{\textbf{IC}2} - \gamma_i  \geq \frac{ I^t_{q}(m') - I^t_{q}(m)}{ r_q^t(m') - r_q^t(m)} ,~~\forall m'>m, \label{ic2}\end{align}
so  we should have $\gamma_i \in [l_{q,m}^\gamma(\mathbf{I}^{t}_q ), h_{q,m}^\gamma(\mathbf{I}^{t}_q )]$, with
\begin{align} &h_{q,m}^\gamma(\mathbf{I}^{t}_q ) = \min\left\{\frac{I^t_{q}(m)- I^t_{q}(m')}{r_q^t(m)- r_q^t(m')}|_{0\leq m'<m}\right\},\\
&l_{q,m}^\gamma(\mathbf{I}^{t}_q ) = \max\left\{  \frac{ I^t_{q}(m') - I^t_{q}(m)}{ r_q^t(m') - r_q^t(m) }|_{m'>m}\right\}. \end{align}
which gives,
\begin{align} &P(E_{q,m}(\mathbf{I}^{t}_q )) = F_{\gamma}^{q}(h_{q,m}^\gamma(\mathbf{I}^{t}_q ) ) - F_{\gamma}^{q}(l_{q,m}^\gamma(\mathbf{I}^{t}_q ) )\label{probqm}\end{align}

However,  due to the absence of any natural ordering, these constraints will render the optimization problem \eqref{profitmax} rather complex. Thus, next, we will impose a design constaint that ensures that {\it local incentive compaibility} is sufficient for decision making, i.e., if the customer prefers mode $m$ over adjacent modes $m+1$ and $m-1$, he/she will also prefer mode $m$ over all modes $m'>m$ and $m'<m$. 

\textbf{Design constraint 2} (Single-Crossing Incentive Profile): we will design the incentive profile such that $\forall m \in \mathcal{M}_q$, the ratio $ \frac{I^t_{q}(m+1)- I^t_{q}(m)}{r_q^t(m+1) - r_q^t(m)}$ is non-increasing, i.e., incentives grow slower than risks for higher modes $m$.

\begin{proposition}If the incentive profile is single-crossing, customer $i$ will pick mode $m\geq 1$ simply iff
\begin{align}  &   \frac{I^t_{q}(m+1)- I^t_{q}(m)}{r_q^t(m+1) - r_q^t(m)}  \leq \gamma_i  \leq \frac{I^t_{q}(m)- I_{q}(m-1)}{r_q^t(m) - r_q^t(m-1)},\label{localinc}
 \end{align}
where the right hand inequality ensures \eqref{ir} and \eqref{ic1}, and the left hand inequality ensures \eqref{ic2}. To keep  expression \eqref{localinc} compact, we use a  dummy mode $m = M_q+1$, with $I_q(M_q+1) = I_q(M_q)$, and $r_q^t(M_q+1) = r_q^t(M_q) + 1$.\end{proposition}

We acknowledge that the single-crossing condition restricts the values that the incentive profile can take into a region that may be suboptimal for the aggregator. However, it considerably lowers the numerical effort to solve \eqref{profitmax}. 

 To illustrate the next steps required to solve the problem, specifically required for our numerical experiments, we assume that the types $\gamma_i$ for cluster $q$ are drawn from a uniform distribution over $[0,\gamma_{\max}^q]$. We  design the incentives such that all of the ratios $\frac{I^t_{q}(m+1)- I^t_{q}(m)}{r^t_q(m+1) - r^t_q(m)}$ fall in the probability space of $\gamma_i$.    This happens if the incentives are positive and non-decreasing with $m$, and that the ratio $I^t_{q}(1)/r_q^t(1)$ is not above $\gamma_{\max}^q$, i.e., someone might pick mode $m=1$. Consequently, imposing the diminishing pay-off constraints for deferrable loads, \eqref{profitmax} is written as,
\begin{align}&\max_{\mathbf{I}^{t}_q } \sum_t N^t_q =  \frac{1}{\gamma_{\max}^q} \sum_t\sum_{m=1}^{M_q}\bigg( U^t_q(m) -  I^t_{q}(m)\bigg)\times\nonumber\\&~~~~~~~~~ \bigg(\frac{I^t_{q}(m)- I^t_{q}(m-1)}{r_q^t(m) - r_q^t(m-1)} -  \frac{I^t_{q}(m+1)- I^t_{q}(m)}{r_q^t(m+1) - r_q^t(m)}\bigg),\nonumber\\
&\mbox{s.t.}~~0 \leq I^{t}_q(m) \leq I^{{t}-1}_q(m+1), ~ 1 \leq m \leq M_q-1, \nonumber\\
&~~~~~  \frac{I^t_{q}(m+1)- I^t_{q}(m)}{r_q^t(m+1) - r_q^t(m)} \leq  \frac{I^t_{q}(m)- I^t_{q}(m-1)}{r_q^t(m) - r_q^t(m-1)}, ~ m \geq 1, \nonumber \\
&~~~~~  I^t_{q}(m) \geq I^t_{q}(m-1) , ~ m \geq 1, \nonumber \\
&~~~~~  \frac{I^t_{q}(1)}{r_q^t(1)} \leq \gamma_{\max}^q, \label{qprg}\end{align}
for which the objective function is quadratic in $\mathbf{I}^{t}_q $, and the constraints are all affine.

The statistics on the risk functions can be obtained by conducting market surveys or using learning techniques. The details are out of the scope of this work.

\subsubsection{Model-free Learning Approach} If there is no single underlying model that characterizes how costumers respond to participation incentives, or no information is available on the private risk functions of customers, model-free online learning techniques can be used to directly learn the probabilities $P(E_{q,m}(\mathbf{I}^{t}_q ))$. These approaches explore different alternatives for the incentive signals $\mathbf{I}^{t}_q $, observe the response of the population, an update their estimate of $P(E_{q,m}(\mathbf{I}^{t}_q ))$ accordingly.  This requires that the event of a customer {\it considering} to join the program be observable to the aggregator, which is not the case in our current design. This issue can be addressed by asking the customers to make an {\it anonymous query} every time they need to look at the incentive menus for a specific cluster.

Next, we study the recruitment utility of different devices for the aggregator.

\section{The Recruitment Utility Functions} 
There are several options for an aggregator to profit from recruiting flexible appliances. Here we only study the possibility of load shifting based on the recruitment of long-duration non interruptible deferrable appliances (e.g., EVs, washer/dryers ) or preheating and precooling of TCLs. 

In order to solve \eqref{profitmax}, the aggregator needs to know beforehand how much utility it could expect from recruiting an appliance in cluster $q$ in mode $m$ at time $t$. We denoted this value by $U_q^t(m)$.  In this Section, we specifically focus on  calculating this payoff for the case of load shifting with non-interruptible deferrable loads and Thermostatically Controlled Loads (TCL), with  the numerical results focusing specifically on Plug-in Hybrid Electric Vehicles (PHEV). Due to limited space, we leave the discussion of interruptible and sheddable loads to future work.


Energy is traded on an hourly basis, and the demand is modeled as constant in hourly intervals in the energy market. The task of ensuring the sub-hourly balance of demand and supply is left to ancillary service providers, which are procured and dispatched by the grid operator to respond quickly (in a matter of minutes or seconds) to variations of the demand. Traditionally, ancillary services are offered by fast ramping generators.

In order to serve its' load, the aggregator needs to purchase a certain amount of energy for every hour $\ell$ from the wholesale energy market, which we denote by $L(\ell)$, and refer to as the \textit{base load}. We assume that the aggregator has access to ex-ante forecasts of the wholesale energy market clearing prices, and we denote the expected value of the price for hour $\ell$ as $\pi^e(\ell)$. We assume the aggregator's load is small enough to not affect the price.  The aggregator can save money in the wholesale energy market if it recruits flexible appliances and shifts their load to hours at which energy is cheaper.

Denote the set of all feasible {\it hourly load traces} of an individual task in cluster $q$ under mode $m$ recruited at time 
$t$ as $L^t_{q,m}(\ell) \in \mathcal{L}_{q,m}^{t}$. To obtain an hourly load profile for an appliance with sub-hourly consumption variations, we simply average out the total consumption of the device within each hour. We denote as  $L^t_{q,0}(\ell)$ the consumption profile of the appliance if it is not recruited by the DLS program and starts its consumption at time $t$. Serving this power to the consumer through buying energy from the wholesale market presents a cost for the aggregator. By having direct control over this appliance, the aggregator  can expect to gain the following payoff by shifting the load away from peak hours:
\begin{eqnarray} U_q^t (m) &=& \sum_{\ell} \pi^e(\ell)L_{q,0}^t(\ell) \\ &-&\min_{L_{q,m}^t(\ell) \in \mathcal{L}_{q,m}^{t}} \sum_{\ell}\pi^e(\ell)L_{q,m}^t(\ell). \label{basicshift} \nonumber \end{eqnarray}

\begin{remark}{\it 
The consumption profile of flexible appliances that do not choose to join the DLS program, i.e., $L_{q,0}^t(\ell)$, is a function of the tariff that the customers are billed on. On a flat tariff, the customers do not need to spend any effort to find the best time at which they should consume electricity, and simply plug in an appliance at the request arrival time.
%
}
\end{remark}

\textbf{Design constraint 3}: Here we assume that for all deferrable loads, the mode index $m \in \mathcal{M}_q$  is directly equivalent  to the amount of  laxity (slack time) that accompanies the request.

\subsection{Deferrable Non-interruptible Appliances}
Here we look at appliances whose consumption profile can merely be shifted in time, but cannot be modified in any way once started. A classic example is that of a washing machine cycle. In this case, the control variable is the {\it activation time} of the task, denoted by $\alpha$ (See Fig. 1).  The set of possible values of $\alpha$ depends on the initial time at which the request arrives, and the laxity that accompanies the request, i.e., $m$.

  Once activated, the appliance consumes a predetermined amount of power, denoted by the time-shifted pulse $g_{q}(j-\alpha)$ for appliances in cluster $q$, with a length of $\Gamma_{q}$ epochs. This pulse  varies on a sub-hourly basis as a function of the time epochs $j$. To relate this with  the variables required to solve the optimization \eqref{basicshift}, we need to map this sub-hourly varying load to an hourly load profile,  resulting in  
\begin{eqnarray}\mathcal{L}_{q,m}^{t} = \left\{ L_{q,m}^t(\ell) = \sum_{j=\ell S}^{(\ell+1)S-1} g_{q}(j-\alpha)| t\leq \alpha\leq t+m\right\}.\nonumber\end{eqnarray}


Notice the summations required to calculate the average hourly consumption of the device from the sub-hourly load. To avoid the inconvenience of going from $g_{q}(j-\alpha)$ to hourly consumption values $L_{q,m}^t(\ell)$ when solving \eqref{basicshift},  we simply expand the expected wholesale energy price vector $\pi^e(\ell)$ to define virtual sub-hourly wholesale prices $\pi^p(j)$,
$$\pi^p(j) = \sum_{\ell} \frac{1}{S} \pi^e(\ell) \Pi(\frac{ j-\ell S}{S} ),$$
where $\Pi(.)$ denotes the unit pulse function between $[0,1)$. With this new definition, we have
\begin{eqnarray}  \sum_{\ell = \lfloor \frac{\alpha}{S} \rfloor}^{\lfloor \frac{\alpha+\Gamma_q}{S} \rfloor}\pi^e(\ell)L_{q,m}^t(\ell) &=& \sum_{\ell = \lfloor \frac{\alpha}{S} \rfloor}^{\lfloor \frac{\alpha+\Gamma_q}{S} \rfloor}\pi^e(\ell)\sum_{j=\ell S}^{(\ell+1)S-1} g_{q}(j-\alpha) \nonumber \\ &=& \sum_{j = \alpha}^{\alpha+ \Gamma_{q}}\pi^p(j) g_{q}(j-\alpha).\end{eqnarray}

Consequently, the expected utility of recruiting an appliance in cluster $q$ under mode $m$ at time $t$ is
\begin{eqnarray} U^t_q(m) &=& \sum_{j = t}^{t + \Gamma_{q}}\pi^p(j) g_{q}(j-t) - \nonumber\\&~&\min_{\alpha} \left[\sum_{j = \alpha}^{\alpha+ \Gamma_{q}}\pi^p(j) g_{q}(j-\alpha)\right] \nonumber\\\mbox{s.t.}&~& t\leq \alpha\leq t+m, \label{scen1}\end{eqnarray}
where the first term refers to the cost incurred by the aggregator if no demand control is exercised, whereas the second term mirrors the lowest possible cost with which the appliance can be served, considering a demand laxity of $m$ time units.

\subsection{Thermostatically Controlled Loads}

Another category of appliances that can help the aggregator save money in the energy market are TCLs, which can use the inherent energy storage property of building thermal mass to shift their load to cheaper hours. A substantial potential for intrahourly load shifting with heating and cooling devices is through preheating and precooling of the air-conditioned space. To preheat a space, the temperature is increased above the comfort band of the user at off-peak hours, before residents/employees arrive at home/work. The preheating should be scheduled such that by the time the residents arrive, the temperature is close to the  highest acceptable temperature in the comfort band. While some commercial building utility managers currently exercise this option during night hours to save on energy, with the integration of renewable resources, the {\it best time and amount} of preheating would be variable on a daily basis and needs coordination with the aggregator. 

We assume that each TCL is equipped with a thermostat that keeps the temperature in a certain comfort band $[x^{\min}_q,x^{\max}_q]$, with $x^{\min}_q$ and $x^{\max}_q$ representing the lower and upper thresholds chosen by the customers in cluster $q$. 
 Here, we adopt a simple first-order model proposed in \cite{model2} to explain the state dynamics of temperature. Let $x_i(j)$ (a representative temperature) characterize the state of the $i$-th heating device; $x_i(j)$ evolves according to the linear stochastic difference equation
\begin{equation}\label{tcldiff}
x_i(j) - x_i(j-1) = -k_q (x_i(j-1) - x_a(j-1)) +W_q b_i(j) + \nu_i(j),
\end{equation}
where we denote by
\begin{itemize}
\item $k_q$ the average loss rate for buildings in cluster $q$;
\item $x_a(j)$ the ambient temperature;
\item $W_q$ the average rate of heat  gain supplied by devices in cluster $q$;
\item $b_i(j)$ the operating state of the device at time $j$ (1 for ''on" or 0 for ''off");
\item  $\nu_i(j)$ a zero mean Wiener noise process.
\end{itemize}

We assume that, once on, the unit consumes a constant power approximately equal to $P_q$ for cluster $q$. Thus, the only variable that needs to be chosen and can affect the aggregator's cost is the times at which to run the unit for preheating. Due to the constant power of the device, the amount of preheating is a direct consequence of this choice. However, we would like to point out that more sophisticated methods for determining the trajectory of the temperature for precooling/preheating exists in the literature, both based on optimal control theory \cite{tclpre1, tclpre2, tclpre3,hasan2012aggregator}, or learning and simulation \cite{tclpre4}. The calculations in the section could be updated to account for any of these techniques.

In order for the temperature to be around $x^{\max}_q$ at the time that the building occupants arrive, we must have previously increased it enough in the preheating period. This preheating process can be done over multiple disjoint time intervals, whenever it is deemed as a cost-effective action.

For appliances recruited at time $t$, preheating can start right away and can continue to the time at which building occupants arrive, which is equal to $t+m$, with the mode index $m$  directly mirroring the time laxity offered for preheating by the customer. Thus, the optimization that needs to be solved to minimize the expected cost for this is
\begin{eqnarray}  C^{\rm preheat}_{q}(m)= &~&\min_{b_i(j)} \left[\sum_{j = t}^{t+m}\pi^p(j)  b_i(j)\right] \nonumber\\\mbox{s.t.}&~&x^{\max}_q - \delta \leq x_q(t+m) \leq x^{\max}_q+\delta,\nonumber\\ &~&b_i(j) \in [0,1] \end{eqnarray}
where $\delta \geq 0$ should be picked such that at least one number in the set $[x^{\max}_q - \delta,x^{\max}_q + \delta]$ lies within the controllability subspace of the dynamics that govern $x_i(t+m)$.  To write these dynamics in closed form, we assume that $x_i( t)  = x_a( t)$, i.e.,  we assume that the unit stays off long enough when the building is vacant, and the temperature $x(t)$  has already reached the ambient temperature $x_a(t)$ when preheatig starts. Then, we can write the temporal evolution of the expected value of  $x_i(j)$ as
\begin{equation}\label{diffx}
x_i(j) = c_i(j) + \sum_{w= t}^{j-1} (1-k_q)^{j-w-1}  W_q b_i(w)
\end{equation}
with $c_i(j) = \sum_{w=t}^{j-1} (1-k_q)^{j-w-1} \Big( x_i(w) (1-k_q) + k_q x_a(w) \Big)$. The $b_i(w)$'s are the decision variables.

In order to see how much benefit direct scheduling can provide to the aggregator, we need to compare the cost to the case when no preheating is exercised. In that case, heating starts at $t+m$, and it will take the TCL $\tau^{up}_q$  units of time to continuously work and get the temperature to $x_q^{\max}$, where
\begin{equation}\label{ji}
\tau^{up}_q \approx \frac{1}{k_q}\ln\bigg(\frac{\frac{W_q}{k_q}} {x^{\max}_q - \frac{W_q}{k_q} - x_a(t+m)} \bigg).
\end{equation}
This assumes that the dynamics of the ambient temperature is much slower than that of the TCL. The first heating cycle right after turning on in the morning will have an expected cost equal to
$$ C^{\rm normal}_q = \sum_{j = t+m}^{t+m+\tau^{up}_q-1 } \pi^p(j) $$
for the aggaregator. Thus, the utility of recruting a TCL in cluster $q$ in mode $m$ at time $t$ is given by:
$$U_q^t (m)  = \max\left\{C^{\rm normal}_q  -   C^{\rm preheat}_{q}(m), 0\right\} $$
Note that providing any DLS service requires that the appliance is submetered. Here this is essential since the overall consumption of the device could be fairly increased and the consumer should not be billed accordingly.

\section{Numerical Case Study Using EV Data}
We simulate the interactions of one aggregator with a population of Plug-in Hybrid EVs arriving at random to receive level-1 battery charge (1.1 kW instantaneous rate). The arrival time, charge duration, and laxity data are taken from real PHEV charge events recorded and studied in \cite{drk1}. This database includes 620 charge events that happen over the length of several months. However, we ignore the dates and treat the  plug-in events as if they happened on the same day. The aggregator interacts with the wholesale energy market. Hourly LMPs for the one day of operation simulated here are taken  from ISO New England's Maine load zone on September 1st, 2013 to noon of September 2nd.

We translate the absolute charge laxity values in \cite{drk1} into {\it risk functions} by tuning the type $\gamma_i$ in \eqref{typeeq} such that it would not be individually rational for the customer to offer laxities that exceed their real departure time, given one sample of the daily incentive profile. For lack of any meaningful alternative, we take $r_q^t(m) = m$. Denote the mode corresponding to the maximum laxity that could be offered while allowing a timely departure for the customer by $m_d$. Then we assume that for each charging event in cluster $q$ at time $t$, given sample incentives $\hat{I}^t_q(m)$, the risk type $\gamma_i$ of the user $i$ is such that  $$ (m_d+1) \gamma_i \geq \hat{I}^t_q(m_d+1).$$

Next, in order to solve the optimization \eqref{qprg}, we fit a uniform distribution on the customer types, resulting in $$\gamma_i \sim U(0,0.0721).$$

One can use the Gershgorin circle theorem to show that the optimization problem \eqref{qprg} is concave (with a negative semidefinite quadratic term). Incentives are designed for half-hour intervals, i.e., $t = 1,\ldots, 48$ for one full day. Remember that the solution to this optimization is sub-optimal due to the single-crossing condition and the uniform prior assumed for the customer risk types $\gamma_i$. 

\subsection{Comparison of Aggregator Payoff for Different Incentive Design Schemes}
\begin{figure}
\centering
\includegraphics[width = \linewidth]{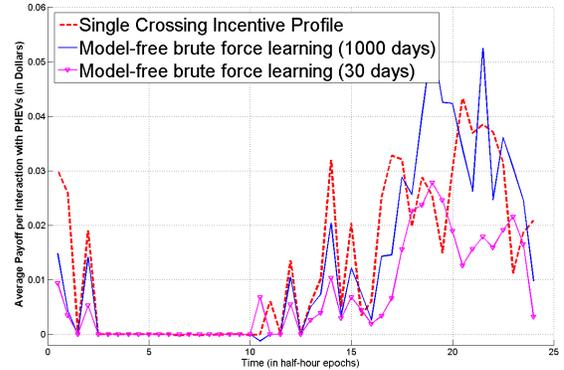}
\caption{Comparison of aggregator payoff when offering incentives designed through different schemes to a PHEV population}
\label{comp3}
\end{figure}

The average payoff ($N^t_q$) that the aggregator receives at different times of the day $t$ when interacting with the above-described customers under the   single-crossing incentives is shown in Fig. \ref{comp3} (dashed red curve).
To study the level of sub-optimality imposed by the single-crossing design constraint, we compare the outcome of the model-based incentive design technique to a  brute-forced learning method, where several scenarios are generated for the daily incentive profile and tested on the population to observe the response. With more trials, the aggregator would find better incentive profiles. Fig. \ref{comp3} compares the aggregator's profit with the single-crossing profile with that of the brute-force learning method after 30 and 1000 days. The reader can see that the learning method is performing better than the single-crossing profile after 1000 days, but not in 30 days.

\subsection{Welfare Effects}
\begin{figure}
\centering
\includegraphics[width = \linewidth]{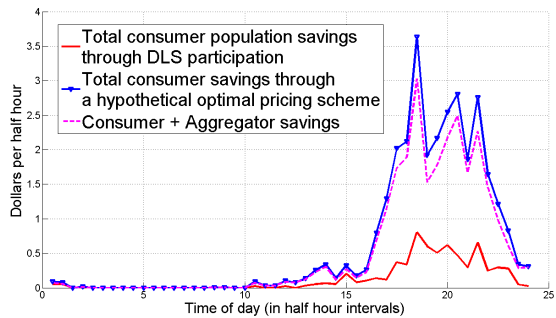}
\caption{Aggregate Consumer savings for the 620 Charging Events}
\label{surp}
\end{figure}

We define the consumer savings through a DSM technique as the expected change in the monetary value of maximum utility across different alternatives for receiving electricity service. As explained before, since the basic service that is being received/provided through any demand management program is always the same as the normal operation mode of the grid today, i.e., there is no change in overall consumption and only the timing of consumption changes, we ignore the utility of receiving the standard service of electricity consumption in these calculations.  The consumer savings  by participating in the DLS program (under brute-force optimized incentives after 1000 days) is equal to $I^t_{q_i}(m_i)- R^t_{i}(m_{i})$. Saving impacts are individually calculated for each
charge request, and the sum of this value over the 620 charging events is shown in Fig. \ref{surp} as a function of time of plug-in.

Now assume that,  instead of going through a  intermediary node like the aggregator, the customer could have, through a  hypothetical demand management scheme,  directly interacted with the electricity grid. Hence, customers could submit their consumption flexibility as bids to the Independent System Operator (ISO) that runs the market, and are billed based on optimal market clearing prices. This is the ultimate goal of dynamic pricing schemes and could realize the most efficient outcome of the market. Alternatively, one could imagine any scheme that   uses the full potential of customers for load shifting, and relays all the wholesale market savings to them. For now, assume that today's regulated flat tariffs  are designed such that customers are billed at the  averaged wholesale cost of the electricity they consume (ignoring aggregator mark up). Then, the consumer savings through the implementation of such dynamic pricing program would have been equal to the recruitment utility, i.e., $U_q^t(m_d)$, assuming that customers would use their maximum possible laxity with no reservation when making individual scheduling decisions in response to a price (no commitment risk function). The value of this quantity is summed up across the population and is compared to customer savings gained from the DLS scheme in Fig. \ref{surp}.

The considerable difference in  consumer savings between the DLS program and that of optimal pricing is due to two factors: 1) the presence of an intermediary for-profit node, i.e., the aggregator; 2) the commitment risk function $R^t_{i}(m_{i})$, preventing the customers from realizing all their load shifting potentials. To observe the welfare effects of the second phenomena, we show the sum of the aggregator and consumer savings under DLS in Fig. \ref{surp}. The reader can observe that the overall decrease in the community's (aggregator + consumers) welfare is minimal. Remember that this decrease in welfare comes with the promise of reliability and controllability, which are the essential attributes that drive electricity market design problems and are required for safe grid operation.

\section{Conclusions and Future Work}
Due to the popularity of direct control mechanisms in the power grid, designing appropriate incentives for customer participation in these programs seems like an inevitable issue. Here we approached this issue as a market design problem to trade flexibility with an aggregator of electricity services.  Extension of this work to look at the possibility of providing ancillary services through collective effort of several appliances, and to study competition between several aggregators that can serve the same  population, is left to future work.

\bibliographystyle{IEEEtran}
\small
\bibliography{New,Science,Science2}
\end{document}